# Self-Confirming Price Prediction Strategies for Simultaneous One-Shot Auctions

**Michael P. Wellman**
Computer Science & Engineering
University of Michigan

**Eric Sodomka**
Computer Science
Brown University

**Amy Greenwald**
Computer Science
Brown University

## Abstract

Bidding in simultaneous auctions is challenging because an agent's value for a good in one auction may depend on the uncertain outcome of other auctions: the so-called *exposure problem*. Given the gap in understanding of general simultaneous auction games, previous works have tackled this problem with heuristic strategies that employ probabilistic price predictions. We define a concept of *self-confirming prices*, and show that within an independent private value model, Bayes-Nash equilibrium can be fully characterized as a profile of optimal price-prediction strategies with self-confirming predictions. We exhibit practical procedures to compute approximately optimal bids given a probabilistic price prediction, and near self-confirming price predictions given a price-prediction strategy. An extensive empirical game-theoretic analysis demonstrates that self-confirming price-prediction strategies are effective in simultaneous auction games with both complementary and substitutable preference structures.

## 1 Introduction

One of the most attractive features of automated trading is the ability to monitor and participate in many markets simultaneously. Compared to human traders, software agents can take in data from multiple sources at very high throughput rates. In principle, software agents can also process massive quantities of information relevant to trading decisions in short time spans. In practice, however, dealing with multiple markets poses one of the greatest strategic challenges for automated trading. When markets interact, a strategy for bidding in one market must consider the implications of and ramifications for what happens in others.

Markets are *interdependent* when an agent's preference for the outcome in one depends on results of others. For instance, when value for one good is increased by obtaining another, the goods are *complements*. Dealing with multiple markets under complementary preferences presents an agent with the classic *exposure problem*: before it can obtain a valuable bundle, the agent must risk getting stuck with a strict subset of the goods, which it may not have wanted at the prevailing prices. Exposure is a potential issue for *substitute* goods as well, as the agent risks obtaining goods it does not want given that it obtains others.

The pitfall of exposure is a primary motivation for combinatorial auctions [Cramton et al., 2005], where the mechanism takes responsibility for allocating goods respecting agents' expressed interdependencies. Combinatorial auctions are often infeasible, however, due to nonexistence of an entity with the authority and capability to coordinate markets of independent origin. Consequently, interdependent markets are inevitable. Nonetheless, there is at present very little fundamental understanding of agent bidding strategies for these markets. Specifically, how should an agent's bidding strategy address the exposure problem?

We address this question in what is arguably the most basic form of interdependent markets: simultaneous one-shot sealed-bid (SimOSSB) auctions. Despite the simplicity of this mechanism and the practical importance of the exposure issue, there is little available guidance in the auction theory literature on the strategic problem of how to bid in SimOSSB auctions. We aim to fill this gap by providing computationally feasible methods for constructing bidding strategies for the SimOSSB-auction environment, which we justify with both theory and evidence from simulation-based analysis. Specifically, we (i) characterize Bayes-Nash equilibria of SimOSSB auctions as best responses to price predictions (§3); (ii) provide bounds on approximate Bayes-Nash equilibria in terms of the accuracy of price predictions and the degree of optimality of responses (§4); (iii) introduce methods to construct bidding strategies that respect the equilibrium form (§5,6); and (iv) demonstrate through a comprehensive empirical game-theoretic analysis the efficacy of these strategies compared to a wide variety of heuristics from the literature (§7).

## 2 Previous Work

Theoretical results about general simultaneous auction games are few and far between. The leading auction theory textbook [Krishna, 2010] treats sequential but not simultaneous auctions, and most of the literature that addresses simultaneity does so only in the context of ascending [Peters and Severinov, 2006] or multi-unit auctions.

In the first work to derive an equilibrium of a simultaneous-auction game, Engelbrecht-Wiggans and Weber [1979] tackle an example with perfect substitutes, where each agent is restricted to bid on at most two items. Their analysis was performed in the large-limit of auctions and agents, and exhibited a mixed equilibrium where the agents diversify their bids even though the items are indistinguishable. Krishna and Rosenthal [1996] studied a second-price setup with two categories of bidders: local bidders who have value for a single item, and global bidders who have superadditive values for multiple items. The authors characterize an equilibrium that is symmetric with respect to the global bidders, and show, somewhat surprisingly, that an increase in the number of bidders often leads to less aggressive bidding. Rosenthal and Wang [1996] tackled a first-price setup, assuming synergies and common values. Szentes and Rosenthal [2003] studied two-bidder auctions with three identical objects and complete information.

Recently, Rabinovich et al. [2011] generalized fictitious play to incomplete information games with finite actions and applied their technique to a class of simultaneous second-price auctions. They computed approximate equilibria in environments with utilities expressible as linear functions over a one-dimensional type space.

Complementing these theoretical treatments, researchers have designed trading strategies applicable to simultaneous auctions, which address the exposure problem through heuristic means. In the Trading Agent Competition (TAC) Travel game [Wellman et al., 2007], agents face an exposure problem for hotels—they must obtain a room each night for the client or the whole trip is infeasible. Experience from TAC and other domains has demonstrated the importance of *price prediction* for bidding in interdependent markets [Wellman et al., 2004]. Given probabilistic predictions of prices across markets, agents can manage exposure risk, choosing bids that trade off the profits and losses of the possible bundles of goods they stand to win.

Greenwald and Boyan [2004] framed the problem of bidding across interdependent markets given probabilistic price predictions. Follow-on work [Greenwald et al., 2009, Wellman et al., 2007] formalized this bidding problem in decision-theoretic terms and established properties of optimal bidding assuming that bids do not affect other-agent behaviors. Further experimental comparison was performed by Greenwald et al. [2010]. These works introduced a taxonomy of heuristic bidding strategies [Wellman, 2011], which we employ here.

Self-confirming price-prediction (SCPP) bidding strategies were first explored in the context of simultaneous ascending auctions (SimAAs) [Cramton, 2005]. For the SimAA environment, SCPP strategies were found to be highly effective at tackling the exposure problem [Wellman et al., 2008]. To our knowledge, no other general price-prediction methods have been proposed for SimOSSB auctions, other than learning from historical observations.

## 3 Price-Prediction Strategies and Equilibrium

We consider a market with $m$ goods, $\mathcal{X} = \{1, \ldots, m\}$, and $n$ agents. Agent $i$'s value for a bundle $X \in 2^{\mathcal{X}}$ is given by $v_i(X)$, where $v_i(X) \in [0, \bar{V}]$. We assume free disposal: if $X \subseteq X'$, then $v_i(X) \leq v_i(X')$. The $m$ goods are allocated to the agents via SimOSSB auctions, one per good. That the mechanism is simultaneous means that each agent $i$ submits a bid vector $\boldsymbol{b}_i = (b_i^1, \ldots, b_i^m) \in \mathbb{R}_+^m$ before a specified closing hour. That the auctions are one-shot means that all auctions compute and report their results upon the closing hour. That the bids are sealed means that agents have no information about the bids of other auction participants until the outcome is revealed.

The *second-price sealed-bid* (SPSB) auction is an OSSB auction in which the winning bidder pays the second-highest bid rather than its own (highest) bid. The environments studied in our empirical game-theoretic analysis below employ the SPSB mechanism. For simplicity of description we focus on SPSB throughout, although our theoretical results hold as well for first-price sealed-bid auctions, or indeed any auction mechanism where the outcome to agent $i$ (whether it gets the good and the price it pays) is a function of $i$'s own bid and the highest other-agent bid.

Our investigation employs the familiar *independent private values* (IPV) model (see, for example, [Krishna, 2010]), where each agent $i$'s values are drawn independently from a probability distribution that is common knowledge. Under IPV, it is a dominant strategy for an agent to bid its true value in a single SPSB auction. This result does not generalize to simultaneous SPSB auctions, however, unless the agent's value over bundles happens to be additive. When agents' values for different goods interdepend (e.g., through complementarity or substitutability), bidding truthfully in simultaneous auctions is not even an option, as the value for an individual good is not well-defined.

To deal effectively with interdependent markets, an agent's bid in each auction must reflect its beliefs about the outcomes of others. We consider beliefs in the form of *predictions* about the prices at which the agent might obtain goods in the respective auctions. Bidding strategies that are explicitly cast as functions of some input price prediction are termed *price-prediction (PP) strategies*.

We denote by $p_j \in \mathbb{R}_+$ a *price* for good $j$. The vector $\boldsymbol{p} = \langle p_1, \ldots, p_m \rangle$ associates a price with each good. We represent price predictions as probability distributions over the joint price space. We use the symbol $\Pi$ for such predictions in the form of cumulative probability distributions,

$$\Pi_{\boldsymbol{p}}(\boldsymbol{q}) = \Pr(\boldsymbol{p} \leq \boldsymbol{q}), \tag{1}$$

where $\boldsymbol{p} \leq \boldsymbol{q}$ holds iff $p_j \leq q_j$ for all $j$. We generally omit the subscript $\boldsymbol{p}$ as understood.

We have previously argued [Wellman et al., 2007] that price prediction is a key element of agent architecture for complex trading environments. Here, we support a stronger claim for the case of SimOSSB auctions: given IPV, *PP strategies are necessary and sufficient for optimal bidding*.

Let $w(\boldsymbol{b}, \boldsymbol{q}) = \{j \mid b^j > q_j\}$ denote the set of goods an agent would win by bidding $\boldsymbol{b}$ when the highest other-agent bids are $\boldsymbol{q}$.[1] Agent $i$'s utility for a bid given others' bids can thus be written as

$$u_i(\boldsymbol{b}, \boldsymbol{q}) = v_i(w(\boldsymbol{b}, \boldsymbol{q})) - \sum_{j \in w(\boldsymbol{b}, \boldsymbol{q})} q_j. \tag{2}$$

**Definition 1** (Optimal PP Bidders). An *optimal PP bidding strategy* $s^*(\Pi)$ submits bids that maximize expected utility given a price prediction $\Pi$,

$$s^*(\Pi) \in \arg\max_{\boldsymbol{b}} \mathbb{E}_{\boldsymbol{q} \sim \Pi}[u_i(\boldsymbol{b}, \boldsymbol{q})]. \tag{3}$$

In games of incomplete information, agent $i$'s strategy produces actions (here, bids) as a function of $i$'s type. Under IPV, knowing other agents' values tells agent $i$ nothing about its own value. The distribution of outcomes given $i$'s bid $\boldsymbol{b}_i$ depends only on $\boldsymbol{b}_i$ and the marginal distribution of other-agent bids. Agent $i$'s expected utility is thus conditionally independent of other-agent valuations given their bids [Wellman et al., 2011]. In particular, $i$'s best response to a profile of other-agent strategies depends only on the distribution of their bids [Rabinovich et al., 2011].

Let $b_{-i}^{k*}$ denote the highest bid submitted for good $k$ by an agent other than $i$. Since agent $i$'s utility depends only on what it wins and what it pays, the distribution of *highest* other-agent bids (i.e., the distribution of $b_{-i}^{k*}$) is a sufficient statistic for the other-agent bid distributions. This distribution can be expressed in the form of a price prediction (1).

Therefore, a best response to other-agent bidding strategies takes the form of an optimal PP bidding strategy (3),

---
[1] In the case of ties, the winner is chosen uniformly from among the high bidders. Our analysis ignores ties, which are rare given our setup of real-valued bids and rich valuation functions. Our theoretical results require that the highest other-agent bid is a sufficient statistic for outcome, which is true as long as one's bid does not tie for best with *two or more* other bidders.

where the input PP is the distribution of highest other-agent bids induced by those other-agent strategies. Since a Bayes-Nash equilibrium (BNE) is a profile of mutual best-response strategies, any profile of optimal PP strategies, where the PP for each equals the distribution of highest bids induced by the other agents' optimal PP strategies, constitutes a BNE. Moreover, any BNE can be characterized as a profile of optimal PP strategies, or mixtures thereof.

**Theorem 1.** *Suppose a SimOSSB auction game, with independent private values, where the outcome of each auction to agent $i$ depends only on $i$'s bid and the highest other-agent bid in that auction. Then the strategy profile $\boldsymbol{s} = (s_i, s_{-i})$ is a Bayes-Nash equilibrium if and only if, for all $i$, $s_i$ is equivalent to an optimal PP bidding strategy with input $\Pi_i(\boldsymbol{q}) = \Pr((b_{-i}^{1*}, \ldots, b_{-i}^{m*}) \leq \boldsymbol{q} \mid s_{-i})$, or equivalent to a mixture of such optimal PP strategies.*

Price predictions that support strategic equilibrium are themselves in a form of equilibrium. The bidding strategies employ price predictions that are actually borne out as correct (i.e., the distributions generated by the strategies are as predicted) assuming that everyone follows the given strategies. This situation can be viewed as a form of *rationalizable conjectural equilibrium* (RCE) [Rubinstein and Wolinsky, 1994], where each agent's conjecture is about the distribution of highest other-agent bids. In this instance, the RCE is also a BNE, since the conjecture provides sufficient information to determine a best response.

An auction game with IPV is *symmetric* if all agents have the same probability distribution over valuations. In earlier work on simultaneous ascending auctions [Wellman et al., 2008], we considered the symmetric IPV case and referred to price predictions in this kind of equilibrium relationship as *self-confirming*. Let $\Gamma$ be an instance of a symmetric IPV SimOSSB auction game, and $s$ a bidding strategy that employs price predictions (whether optimally or not).

**Definition 2** (Self-Confirming Price Prediction (SCPP)). The prediction $\Pi$ is *self-confirming* for PP strategy $s$ in $\Gamma$ iff $\Pi$ is equal to the distribution of the highest other-agent prices $(b_{-i}^{1*}, \ldots, b_{-i}^{m*})$ when all agents play $s(\Pi)$.

The following theorem specializes Theorem 1 for the symmetric case, employing the language of SCPPs.

**Theorem 2.** *In symmetric IPV SimOSSB auctions, a symmetric pure BNE comprises optimal PP bidders employing self-confirming price predictions. Hence, existence of pure symmetric BNE in such games entails existence of self-confirming price predictions.*

In summary, for SimOSSB auctions under IPV, we can restrict attention to optimal PP strategies employing price predictions that are in equilibrium with one another (which for the symmetric case, means SCPPs). In the remainder of this paper, we demonstrate that PP strategies are amenable to effective approximation, are a convenient abstraction on which to design and implement trading strategies, and exhibit a high degree of robustness across environments.

## 4 Approximate Price Prediction

We have shown that an optimal PP strategy is a best response to the strategies of other agents if price predictions $\Pi$ exactly reflect the other-agent bid distributions. Since it is unrealistic to expect perfect price prediction, we examine the consequences of employing PP strategies that use inaccurate price predictions $\Pi'$.

We quantify the inaccuracy of $\Pi'$ in two ways. The first is a multivariate form of the *Kolmogorov-Smirnov (KS) statistic*: $KS(\Pi, \Pi') \equiv \sup_{\boldsymbol{q}} |\Pi(\boldsymbol{q}) - \Pi'(\boldsymbol{q})|$. Second, we define the *bundle probability distance*, $BP(\Pi, \Pi', \boldsymbol{b})$, with respect to a bid $\boldsymbol{b}$: $\sum_{X \subseteq \mathcal{X}} |\Pr_{\boldsymbol{q} \sim \Pi}(w(\boldsymbol{b}, \boldsymbol{q}) = X) - \Pr_{\boldsymbol{q} \sim \Pi'}(w(\boldsymbol{b}, \boldsymbol{q}) = X)|/2$.

We first bound the difference between perceived and actual expected utility when incorrectly using $\Pi'$ instead of $\Pi$. Let us denote the payment for $\boldsymbol{b}$ given highest other-agent bids $\boldsymbol{q}$ by $\psi(\boldsymbol{b}, \boldsymbol{q})$, so that for SPSB we have $\psi(\boldsymbol{b}, \boldsymbol{q}) = \sum_{j \in w(\boldsymbol{b}, \boldsymbol{q})} q_j$. Overall expected utility is the difference between expected value of winnings and payment:

$$\mathbb{E}_{\boldsymbol{q}}[u_i(\boldsymbol{b}, \boldsymbol{q})] = \mathbb{E}_{\boldsymbol{q}}[v_i(w(\boldsymbol{b}, \boldsymbol{q}))] - \mathbb{E}_{\boldsymbol{q}}[\psi(\boldsymbol{b}, \boldsymbol{q})]. \quad (4)$$

To bound the difference between perceived and actual expected utility, we separately consider winnings and payment. The following bounds the amount a bidder can underestimate its expected payment by using $\Pi'$.

**Lemma 3.** *Let $\delta_{KS} = KS(\Pi, \Pi')$ and $\|\boldsymbol{b}\|_1 \equiv \sum_{j=1}^m b^j$. Then for all $\boldsymbol{b}$,*

$$\mathbb{E}_{\boldsymbol{q} \sim \Pi}[\psi(\boldsymbol{b}, \boldsymbol{q})] \leq \mathbb{E}_{\boldsymbol{q} \sim \Pi'}[\psi(\boldsymbol{b}, \boldsymbol{q})] + 2\delta_{KS}(\|\boldsymbol{b}\|_1), \quad (5)$$

*Proof.* See Appendix A.1 in the online supplement. □

We similarly bound the amount a bidder can overestimate its expected value of winnings by using $\Pi'$. A variant distribution $\Pi$ can degrade expected value of winnings only by decreasing the probability of winning valuable bundles. By constraining $BP$ distance, we can ensure, for any set of bundles, that the total probability of winning a bundle from that set at $\boldsymbol{b}$ decreases by at most $\delta_{BP}$. This means that the expected value of winnings can suffer by at most $\delta_{BP}\bar{V}$.

**Lemma 4.** *Let $\delta_{BP} = BP(\Pi, \Pi', \boldsymbol{b})$. Then*

$$\mathbb{E}_{\boldsymbol{q} \sim \Pi}[v_i(w(\boldsymbol{b}, \boldsymbol{q}))] \geq \mathbb{E}_{\boldsymbol{q} \sim \Pi'}[v_i(w(\boldsymbol{b}, \boldsymbol{q}))] - \delta_{BP}\bar{V}. \quad (6)$$

Combining the lemmas, we have the following bound.

**Theorem 5.** *Let $\delta_{KS} = KS(\Pi, \Pi')$ and $\delta_{BP} = BP(\Pi, \Pi', \boldsymbol{b})$. Then for all $i$,*

$$\mathbb{E}_{\boldsymbol{q} \sim \Pi}[u_i(\boldsymbol{b}, \boldsymbol{q})] \geq \mathbb{E}_{\boldsymbol{q} \sim \Pi'}[u_i(\boldsymbol{b}, \boldsymbol{q})] - \delta_{BP}\bar{V} - 2\delta_{KS}\|\boldsymbol{b}\|_1.$$

We can use these bounds to limit how far an optimal PP strategy with inaccurate price predictions $\Pi'$ can be from equilibrium. Let us denote by $\bar{b}$ the maximum payment, that is, the $L_1$-norm of the greatest possible bid vector. The value of $\bar{b}$ is bounded above by $m\bar{V}$ for any rational bidding strategy under any valuation distribution, but typically it will be far less than that.

**Theorem 6.** *Suppose that for all agents $i$, strategy $\hat{s}_i$ is a best response to other-agent highest-bid distribution $\hat{\Pi}_{-i}$, and that $\Pi_{\hat{s}_{-i}}$ is the other-agent bid distribution actually induced by $\hat{s}_{-i}$. If for all $i$, $KS(\hat{\Pi}_{-i}, \Pi_{\hat{s}_{-i}}) \leq \delta_{KS}$, and for all $\boldsymbol{b}$, $BP(\hat{\Pi}_{-i}, \Pi_{\hat{s}_{-i}}, \boldsymbol{b}) \leq \delta_{BP}$, then $\hat{\boldsymbol{s}}$ constitutes an $\epsilon$-Bayes-Nash equilibrium, for $\epsilon = 2\delta_{BP}\bar{V} + 4\delta_{KS}\bar{b}$.*

*Proof.* See online Appendix A.2. □

## 5 Heuristic PP Bidding Strategies

Having shown that optimal PP bidding strategies are theoretically ideal in that they comprise a BNE, we turn our attention to practical PP bidding strategies. Building on prior work [Wellman et al., 2007, Chapter 5], we explore a broad range of heuristic bidding strategies. The most salient of these are described here.

### 5.1 Marginal Values and Optimal Bundles

Interdependence dictates that the value of any individual good must be assessed relative to a bundle of goods. This idea is captured by the notion of *marginal value*.

**Definition 3** (Marginal Value). Agent $i$'s marginal value, $\mu_i(x, X)$, for good $x$ with respect to a fixed bundle of other goods $X$ is given by: $\mu_i(x, X) \equiv v_i(X \cup \{x\}) - v_i(X)$.

Given a fixed vector of prices, $\boldsymbol{p} = \langle p_1, \ldots, p_m \rangle$, let $\sigma_i(X, \boldsymbol{p})$ denote agent $i$'s surplus from obtaining the set of goods $X$ at those prices:

$$\sigma_i(X, \boldsymbol{p}) \equiv v_i(X) - \sum_{j|x_j \in X} p_j. \quad (7)$$

**Definition 4** (Acquisition [Boyan and Greenwald, 2001]). Given price vector $\boldsymbol{p}$, the *acquisition problem* selects an optimal bundle of goods to acquire: $X^* = \text{ACQ}_i(\boldsymbol{p}) \equiv \arg\max_{X \subseteq \mathcal{X}} \sigma_i(X, \boldsymbol{p})$.

Faced with perfect point price predictions, an optimal bidding strategy would be to compute $X^* = \text{ACQ}_i(\boldsymbol{p})$ and then to buy precisely those goods in $X^*$. By definition, this strategy yields the optimal surplus at these prices: $\sigma_i^*(\boldsymbol{p}) \equiv \sigma_i(\text{ACQ}_i(\boldsymbol{p}), \boldsymbol{p})$.

To assess goods with respect to (typically imperfect) point price predictions, we extend the concept of marginal value. Let $\boldsymbol{p}[p_j \leftarrow q]$ be a version of the price vector $\boldsymbol{p}$ with the $j$th element revised as indicated: $\boldsymbol{p}[p_j \leftarrow q] = \langle p_1, \ldots, p_{j-1}, q, p_{j+1}, \ldots, p_m \rangle$.

**Definition 5** (Marginal Value at Prices). Agent $i$'s *marginal value* $\mu_i(x_j, \boldsymbol{p})$ for good $x_j$ at prices $\boldsymbol{p}$ is given by: $\mu_i(x_j, \boldsymbol{p}) \equiv \sigma_i^*(\boldsymbol{p}[p_j \leftarrow 0]) - \sigma_i^*(\boldsymbol{p}[p_j \leftarrow \infty])$.

Here, $\sigma_i^*(\boldsymbol{p}[p_j \leftarrow 0])$ represents the optimal surplus at the given prices, assuming good $x_j$ is free. Similarly, $\sigma_i^*(\boldsymbol{p}[p_j \leftarrow \infty])$ represents the optimal surplus at the given prices, if $x_j$ were unavailable. The difference is precisely the marginal value of good $x_j$ with respect to the prices of other goods. Note that Definition 5 generalizes Definition 3, under the interpretation that goods in $X$ have zero price, and all other goods have infinite price.

**StraightMV** The heuristic strategy StraightMV employs this concept directly, and simply bids marginal value for all goods: $\text{StraightMV}_j = \mu(x_j, \boldsymbol{p})$.

## 5.2 Bidding with Price Distributions

A point price estimate fails to convey the uncertainty inherent in future prices. Probability distributions over prices provide a more general representation, expressing degrees of belief over the possible prices that might obtain.

The *expected value method* [Birge and Louveaux, 1997] approximates a stochastic optimization by collapsing probability distributions into point estimates through expectation. Let $\hat{\boldsymbol{p}}_\Pi = \langle \hat{p}_1, \ldots, \hat{p}_m \rangle$, where $\hat{p}_j = \mathbb{E}_{\boldsymbol{p} \sim \Pi}[p_j]$ is the expectation of $p_j$ under given price prediction $\Pi$.

Any bidding strategy defined for point price predictions can be adapted to take as input distribution price predictions through this expected value method, simply by using $\hat{\boldsymbol{p}}_\Pi$ for the point price prediction. For example, we define StraightMU as the strategy that takes as input a distribution prediction, and bids as StraightMV at the mean prices. "MU" stands for "marginal utility", but its usage here is simply to distinguish the name from its corresponding point strategy ending with "MV".

We implemented two approaches to calculating $\hat{\boldsymbol{p}}_\Pi$. The first samples from $\Pi$, and the second computes the exact expectation of a piecewise approximation of $\Pi$. For the sampling method, accuracy depends on the number of samples $k$ drawn; thus we indicate a version of the strategy by appending "$k$" to the name. For example, StraightMU8 takes the mean of eight samples from $\Pi$, and employs StraightMV with that average price vector as input.

**AverageMU** Whereas StraightMU bids the marginal value of the expected price, the PP strategy AverageMU bids the expected marginal value: $\text{AverageMU}_j = \mathbb{E}_{\boldsymbol{p} \sim \Pi}[\mu(x_j, \boldsymbol{p})]$. Our implementation samples from the price distribution, calculates marginal values for each sample, and averages the results.

## 5.3 Explicit Optimization

Finally, we consider strategies that explicitly attempt to optimize bids given a distribution price prediction, as do the optimal PP bidders introduced above (Definition 1).

**BidEval** One heuristic optimization approach is to generate candidate bid vectors, and evaluate them according to the price prediction. The BidEval strategy uses other bidding heuristics to propose candidates, and estimates each candidate's performance by computing its expected utility given the price prediction. It then selects the candidate bid vector with the greatest expected utility. There are many variations of BidEval, defined by:

- *the method used to generate candidates.* When this method is a named bidding strategy, we indicate this in parentheses; for instance, BidEval(SMU8) generates candidates using StraightMU8. Strategy BidEvalMix employs a mix of previously described bidding strategies to generate candidates.
- *number of candidates generated.* Methods based on sampling naturally produce a diverse set of candidates. For example, each invocation of StraightMU8 employs a new draw of eight samples from $\Pi$ to estimate $\hat{\boldsymbol{p}}_\Pi$, thus we generally obtain different bids.
- whether the candidates are evaluated by exact computation on a piecewise version of $\Pi$, or by sampling. If by sampling, then how many samples are used.

**LocalBid** The LocalBid strategy (see Alg. 1) employs a local search method in pursuit of optimal bids. Starting with an initial bid vector proposed by another heuristic strategy, LocalBid makes incremental improvements to that bid vector for a configurable number of iterations. Those incremental improvements are made good-by-good, treating all other goods' bids as fixed. Assuming bids for all goods except $j$ are fixed, the agent effectively faces a single auction for good $j$, with the winnings for other goods determined probabilistically. For a single SPSB auction, it is a dominant strategy to bid one's expected marginal value for good $j$, which is given by $\mathbb{E}_{\boldsymbol{p} \sim \Pi}[v(w(\boldsymbol{b}, \boldsymbol{p}) \cup \{j\}) - v(w(\boldsymbol{b}, \boldsymbol{p}) \setminus \{j\})]$, under these circumstances.

---
**Algorithm 1** LocalBid
**Input:** prediction $\Pi$, heuristic strategy $s$, #iterations $K$
**Output:** bid vector $\boldsymbol{b}$
  Initialize $\boldsymbol{b} \leftarrow s(\Pi)$
  **for** $k = 1$ to $K$ **do**
    **for** $j = 1$ to $m$ **do**
      $b^j \leftarrow \mathbb{E}_{\boldsymbol{p} \sim \Pi}[v(w(\boldsymbol{b}, \boldsymbol{p}) \cup \{j\}) - v(w(\boldsymbol{b}, \boldsymbol{p}) \setminus \{j\})]$
  return $\boldsymbol{b}$

---

LocalBid is an iterative improvement algorithm: the expected utility of $\boldsymbol{b}$ is nondecreasing with each update. Further, if LocalBid converges, it returns a bid vector that is

*consistent* (unlike AverageMU), in the sense that each element of the vector is the average marginal value for its corresponding good given the rest of the bid.

In an empirical comparison of bid optimization algorithms, we found that the LocalBid method was highly effective. In one representative environment, LocalBid achieved 98.8% optimality (see Figure 1), whereas a version of BidEval produced bids with expected value 94.5% of optimal. In that same environment, the bids generated by AverageMU64 were 66.1% as profitable as the optimal bid.

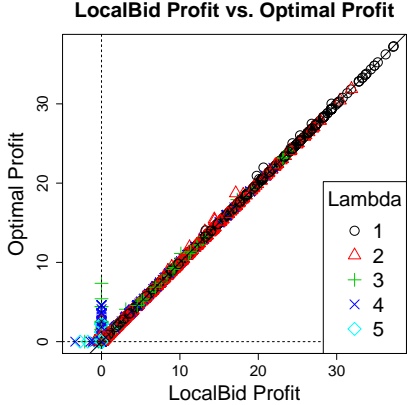

Figure 1: Expected profit for LocalBid versus optimal, for 5000 sample valuations in SimSPSB environment U[5,5]. Points above the diagonal are suboptimal instances.

## 6 Self-Confirming Price Predictions

Now that we have a suite of strategies that employ price predictions, we turn to the question of how to generate such predictions. We propose here to employ *self-confirming price predictions* (SCPPs) (Definition 2), originally introduced and evaluated in the context of simultaneous ascending auctions [Osepayshvili et al., 2005]. A high-level description of our iterative procedure for computing self-confirming price predictions is shown in Alg. 2.

---
**Algorithm 2** Self-Confirming Price Search
---
**Input:** PP strategy PP, parameters $F^0, L, G, \kappa_t, \tau$
**Output:** price prediction $F$
  Initialize $F \leftarrow F^0$
  **for** $t = 1$ to $L$ **do**
    $F' \leftarrow$ outcome of $G$ runs with all playing $\mathsf{PP}(F)$
    **if** $KS_{marg}(F, F') < \tau$ **then**
      **return** $F$
    $F \leftarrow \kappa_t F' + (1 - \kappa_t) F$
  **return** $F$
---

We extend the flexibility of the existing SCPP-derivation procedure by considering two interpretations of the "outcome" prices resulting from each instance of the game. In one, the resulting price is the actual transaction price of the good (price), as in the earlier SimAA study. In the second, we take the highest other-agent bid (HB) as the result.

To measure the difference between distributions at successive iterations, we adopt the Kolmogorov-Smirnov statistic. Since we maintain our prediction in terms of marginal distributions, our comparison takes the maximum of the KS statistic separately for each good: $KS_{marg}(F, F') \equiv \max_j KS(F_j, F_j')$.

Our initial prediction $F^0$ considers all integer prices in the feasible range equally likely. At each iteration, we run $G$ instances of game $\Gamma$, with all agents playing the distribution-prediction strategy $\mathsf{PP}(F)$. We tally the prices resulting from each instance, and update the price prediction as a weighted average of this tally $F'$ and the previous prediction $F$. When $KS_{marg}(F, F')$ falls below threshold, we halt and return $F$. Or, if this distance never falls below threshold, then the procedure terminates and returns the result after $L$ iterations.

Although not guaranteed to converge, we found that for the heuristics of the previous section the iterative procedure generally produced approximately marginally self-confirming distribution predictions. For example, in one environment with a range of strategies we ran the procedure with $G = 10^6$ and $L = 100$, achieving an accuracy $KS_{marg} < 0.01$ in all cases. Due to space constraints, we relegate a full description and evaluation of this procedure to the extended version of this paper.

## 7 Empirical Game-Theoretic Analysis

The heuristic strategies introduced in §5 represent plausible but not generally optimal approaches to bidding in simultaneous auctions. Even the strategies based on explicit optimization (§5.3) fall short of ideal due to inaccuracy in price prediction and non-exhaustive search of bid candidates. To evaluate the performance of these strategies, we conducted an extensive computational study, simulating thousands of strategy profiles—millions of times each—in five different simultaneous SPSB environments. Analysis of the game model induced from simulation data provides evidence for the efficacy and robustness of approximately optimal PP strategies across these environments.

### 7.1 Approach

The methodology of applying game-theoretic reasoning to simulation-induced game models is called *empirical game-theoretic analysis* (EGTA) [Wellman, 2006]. In EGTA, we simulate profiles of an enumerated strategy set playing a game, and estimate a normal-form game from the observed payoffs. The result is a simulation-induced game model, called the *empirical game*. By applying standard game-theoretic solution concepts to the empirical game, we can draw conclusions about the strategic properties of the strategies and profiles evaluated. Although the strategy space in the empirical game is necessarily a severely restricted subset of the original, by including a broad set of strategies representing leading ideas from the literature, we can produce relevant evidence bearing on the relative quality of heuristic strategies in the simulated environments.

Our EGTA study of SimSPSB followed these steps.

1. Define an environment: numbers of goods and agents, and valuation distributions.
2. Specify a set of heuristic strategies. For PP strategies, this includes deriving self-confirming distribution price predictions to be input to these strategies, based on the environment defined in Step 1. The full set of strategies included in our EGTA study is described in online Appendix B.
3. Simulate select profiles among these strategies, sampling from the valuation distributions for each simulation instance (at least one million per profile, most profiles two million or more). Calculate mean payoffs for each strategy in each profile.
4. Analyze the empirical game defined by these mean payoffs to identify Nash equilibria, dominance relationships, regret values, and other analytic constructs.

In actuality, Steps 2–4 were applied in an iterative and interleaved manner, with intermediate analysis results informing the selection of strategies to explore and profiles to sample. The exploration and sample selection were guided manually, generally driven by the objective of confirming or refuting equilibrium candidates among the profiles already evaluated. The process for each environment was terminated when all of the following conditions were met: (1) a broadly representative set of heuristic strategies were covered, (2) all symmetric mixed profiles evaluated were either confirmed or refuted as equilibria, and (3) all strategies showing relative success in at least one environment were evaluated against the equilibria in all other environments. Overall, the analysis commanded some tens of CPU-years over a roughly six-month period.

## 7.2 Environments

We evaluated five simultaneous SPSB environments, involving 3–8 agents bidding on 5 or 6 goods. The environments span two qualitatively different valuation distributions, from highly complementary to highly substitutable. Both of these assume IPV and symmetry, so that each agent receives a private valuation drawn independently from the same distribution.

### 7.2.1 Scheduling Valuations

The first valuation distribution we employ in this study is based on a model of *market-based scheduling* [Reeves et al., 2005]. Goods represent time slots of availability for some resource: for example, a machine, a meeting room, a vehicle, or a skilled laborer. Agents have tasks, which require this resource for some duration of time to complete.

Specifically, the goods $\mathcal{X} = \{x_1, \ldots, x_m\}$ comprise a set of $m$ time slots available to be scheduled. Agent $i$'s task requires $\lambda_i$ time slots to accomplish, and the agent values a set of time slots according to when they enable completion of the task. If agent $i$ acquires $\lambda_i$ time slots by time $t$, it obtains value $V_i^t$. Completion value with respect to time is a nonincreasing function: for all $i$, if $t < t'$ then $V_i^t \geq V_i^{t'}$. If it fails to obtain $\lambda_i$ slots, the agent accrues no value ($V_i^\infty = 0$). Let $X \subseteq \mathcal{X}$ denote a set of slots. The expression $|\{x_j \in X \mid j \leq t\}|$ represents the number of these that are for time $t$ or earlier. The overall valuation function for agent $i$ is $v_i(X) = V_i^{T(X, \lambda_i)}$, where

$$T(X, \lambda) = \min\left(\{t \text{ s.t. } |\{x_j \in X \mid j \leq t\}| \geq \lambda\} \cup \{\infty\}\right)$$

is the earliest time by which $X$ contains at least $\lambda$ slots.

For each agent, a task length $\lambda_i$ is drawn uniformly over the integers $\{1, \ldots, m\}$. Values associated with task completion times are drawn uniformly over $\{1, \ldots, 50\}$, then pruned to impose monotonicity [Reeves et al., 2005]. The valuations induced by this scheduling scenario exhibit strong complementarity among goods. When $\lambda > 1$, the agent gets no value at all for goods in a bundle of fewer than $\lambda$. On the other hand, there is some degree of substitutability across goods when there may be multiple ways of acquiring a bundle of the required size.

We denote environments using this valuation by $U[m, n]$, with $m$ and $n$ the numbers of goods and agents, and "U" indicating the uniform distribution over task lengths.

### 7.2.2 Homogeneous-Good Valuations

The second valuation distribution expresses the polar opposite of complementarity: goods are perfect substitutes, in that agents cannot distinguish one from another. Agents' marginal values for units of this good are weakly decreasing. Specifically, valuation is a function of the number of goods obtained, constructed as follows. Agent $i$'s value for obtaining exactly one good, $v_i(\{1\})$, is drawn uniformly over $\{0, \ldots, 127\}$. Its value for obtaining two, $v_i(\{1, 2\})$, is then drawn from $\{v_i(\{1\}), v_i(\{1\}) + 1, \ldots, 2v_i(\{1\})\}$. In other words, its marginal value for the second good is uniform over $\{0, \ldots, v_i(\{1\})\}$. Subsequent marginal values are similarly constrained not to increase. Its marginal value for the $k$th good is uniform over $\{0, \ldots, v_i(\{1, \ldots, k-1\}) - v_i(\{1, \ldots, k-2\})\}$.

We denote environments using this valuation by $H[m, n]$,

with "H" an indicator for homogeneity.

## 7.3 Regret

We evaluate the stability of a strategy profile by *regret*, the maximal gain a player could achieve by deviating from the profile. Formally, let $\Gamma = \{n, S, u(\cdot)\}$ be a symmetric normal-form game with $n$ players, strategy space $S$ (the same for each player, since the game is symmetric), and payoff function $u : S \times S^{n-1} \to \mathbb{R}$. The expression $u(s_i, s_{-i})$ represents the payoff to playing strategy $s_i$ in a profile where the other players play strategies $s_{-i} \in S^{n-1}$.

**Definition 6** (Regret). The *regret* $\epsilon(\boldsymbol{s})$ of a strategy profile $\boldsymbol{s} = (s_1, \ldots, s_n)$ is given by

$$\epsilon(\boldsymbol{s}) = \max_i \max_{s'_i \in S} \left( u(s'_i, s_{-i}) - u(s_i, s_{-i}) \right).$$

A Nash equilibrium profile has zero regret, and more generally regret provides a measure of approximation to Nash equilibrium. Using this regret definition, profile $\boldsymbol{s}$ is an $\epsilon(\boldsymbol{s})$-Nash equilibrium.

Regret is a property of profiles. Evaluation of a particular strategy is inherently relative to a context of strategies played by other agents. Jordan et al. [2007] proposed ranking strategies according to their performance when other agents are playing an equilibrium.

**Definition 7** (NE regret [Jordan, 2010]). Let $\boldsymbol{s}^{NE}$ be a Nash equilibrium of game $\Gamma$. The regret of strategy $s_i \in S$ relative to $\boldsymbol{s}^{NE}$, $u(s_i^{NE}, s_{-i}^{NE}) - u(s_i, s_{-i}^{NE})$, is an *NE regret* of $s_i$ in $\Gamma$.

NE regret represents the loss experienced by an agent for deviating to a specified strategy from a Nash equilibrium of a game. The rationale for this measure comes from the judgment that all else equal, Nash equilibria provide a compelling strategic context for evaluating a given strategy. For games with multiple NE, a given strategy may have multiple NE-regret values.

## 7.4 Results

Table 1 summarizes the extent of simulation coverage of the five SPSB environments investigated.[2] The empirical games comprise 600–14,000 profiles, over 29–34 strategies. Evaluated profiles constitute a small fraction (as little as 0.03%) of the entire profile space over these strate-gies. Nevertheless, these are sufficient to confirm symmetric Nash equilibria for each game. Whereas it is impossible to rule out additional equilibria without exhaustive evaluation, we have either confirmed or refuted every evaluated symmetric mixed profile (i.e., every subset of strategies for which all profiles are evaluated). Since this includes every strategy in conjunction with those most effective in other contexts, we doubt that there are any other small-support symmetric equilibria, and expect few if any alternative symmetric equilibria among the explored strategies.

---

[2]The extended version includes an online data supplement with full payoff data for the empirical games. The results we report here subsume those of a preliminary study [Yoon and Wellman, 2011]. That study set the groundwork for the current investigation by developing the infrastructure for simulation and derivation of self-confirming price predictions, and tuning parameters (e.g., number of evaluating samples) for several of the strategies. However, the preliminary results reflected sparser coverage of relevant profiles and a weaker overall set of strategy candidates.

Table 1: Strategies and profiles simulated for the environments addressed in our EGTA study.

| Environment | # Strategies | # Profiles | % Profiles |
|---|---|---|---|
| $U[6,4]$ | 34 | 1165 | 1.76 |
| $U[5,5]$ | 30 | 5219 | 1.88 |
| $U[5,8]$ | 29 | 9096 | 0.03 |
| $H[5,3]$ | 32 | 608 | 10.16 |
| $H[5,5]$ | 34 | 13197 | 2.62 |

As it happens, our process identified exactly one symmetric mixed-strategy NE in each game. Of the 44 distinct strategies explored across environments, only seven were supported in equilibrium in any environment. Variants of SCLocalBid, a strategy that explicitly optimizes with respect to self-confirming prices, predominate in equilibrium in four out of five environments. LocalBid and the BidEvalMix strategies also explicitly optimize, but with respect to price predictions that are self-confirming for different strategies (see Appendix B). AverageMU64_HB performs remarkably well in the "U" environments, and is the sole non-optimizing heuristic to appear in equilibria.

Whether a strategy is in equilibrium or not is a crude binary classification of merit. We measure relative degrees of effectiveness by NE regret (Definition 7), as reported in Table 2. The table depicts the NE-regret values for 12 top strategies: all those ranked fourth or better in at least one environment. The values are indicated on a horizontal scale for each game environment, ranging from zero (indicating the strategy is in the support of equilibrium) to the highest NE regret value among the listed strategies. We also verified via a bootstrap technique that the results are statistically robust: for each identified equilibrium the one-tailed 90% confidence bound on regret is at the far left end (1% of the range) of the NE regret scale shown.

These results provide solid support for the SCLocalBid strategy. For the one environment it fails to participate in equilibrium, its NE regret is still quite low, thus we find it to be a strong all-around strategy. This situation contrasts starkly with prior findings for bidding in simultaneous ascending auctions [Wellman et al., 2008], where the best strategies for complementary (scheduling valuation) environments were awful in substitutes (homogeneous good) environments, and vice versa. The fact that SCLocalBid performs so well aligns with our key theoretical finding, in support of optimal PP bidders with self-confirming price predictions. As the LocalBid search method is most effective in optimizing bids, it is consistent with our theory to find that both examples of strategies in this class are leaders among explicit optimizing strategies.

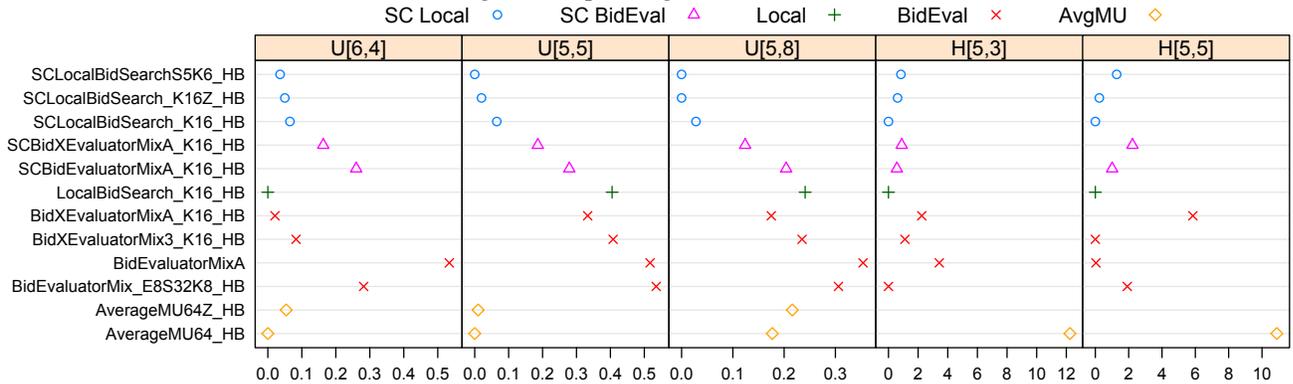

Table 2: NE regret for top strategies across five environments.

All the remaining top strategies are in the BidEval class (also explicit optimizers), except for AverageMU64. In contrast to the others, however, AverageMU64's quality is limited to one of the valuation distributions—the strategy performs poorly in homogeneous-good environments. This observation is consistent with the results reported by Boyan and Greenwald [2001], where an example environment with perfect substitutes was contrived to demonstrate the shortcomings of marginal-utility-based bidding. Given such examples, it is perhaps unsurprising that the heuristic strategies based on marginal value have a difficult time competing with explicit optimizers. If anything it is the observed success of AverageMU64 that is striking, but this outcome is consistent with past experimental results in an environment that exhibits substantial complementarity [Stone et al., 2003]. Still, compared to optimal PP bidders, AverageMU lacks robustness across valuation classes.

Finally, all the top strategies but one employ the highest-bid (HB) statistic in deriving self-confirming price distributions, as opposed to the actual transaction price. This, too, is aligned with what the theory would dictate. The lone exception in Table 2 is BidEvaluatorMixA, which performed impressively in one of the homogenous-good environments but not so well in the rest.

## 8  Conclusion

Our theoretical and experimental findings point to two key ingredients for developing effective bidding strategies for SimOSSB auctions with independent private values. The first is an algorithm for computing approximately optimal bid vectors given a probabilistic price prediction. We have found that a simple local search approach (LocalBid) achieves a high fraction of optimality, and that this translates into superior performance in strategic simulations (§7). The second ingredient is a method for computing self-confirming price predictions for a given price-prediction bidding strategy and a specification of a simultaneous-auction environment. We have found a simple iterative estimation procedure (§6) to be effective at finding price distributions that are approximately marginally self-confirming for a range of strategies and environments.

Our theoretical results say that if these ingredients can achieve their tasks perfectly, we have a solution (i.e., a Bayes-Nash equilibrium) to the corresponding simultaneous auction game. Approximations to the ideal in these ingredients yield approximate game solutions. Our computational experiments indicate that following this approach produces results that are as good or better than any other general method proposed for bidding in SimOSSB auctions. The evidence takes the form of a comprehensive empirical game-theoretic analysis, covering both complementary and substitutable valuation classes and a broad swath of heuristic strategies from the literature.

There is still room for improvement in our proposed methods, as well as opportunity to subject our conclusions to further empirical scrutiny. More sophisticated stochastic search techniques may improve upon our best bid optimization algorithms, and allow them to scale to larger environments with more complex valuations. Similarly, we do not consider our simple iterative method to be a last word on finding self-confirming price distributions. In particular, we expect that substantial improvement could be obtained by accounting for some joint dependencies in price predictions. Finally, further testing against alternative proposals or in alternative environments would go some way to bolstering or refuting our positive conclusions.